\documentclass[reprint,superscriptaddress,amsmath,amssymb,aps,showkeys
]{revtex4-2}

\usepackage{graphicx,xcolor}
\usepackage{dcolumn}
\usepackage{bm}

\newcommand{\av}[1]{\langle #1 \rangle}

\newcommand{\degree}{^\circ}

\DeclareMathOperator\Cor{Cor}

\begin{document}

\preprint{APS/123-QED}

\title{Mixing Property of Symmetrical Polygonal Billiards}

\normalsize

\author{R. B. do Carmo}
\thanks{Corresponding author: ricardo.carmo@ifal.edu.br}
\affiliation{Instituto Federal de Alagoas, Piranhas, AL 57460-000, Brazil}

\author{T. Ara\'ujo Lima}
\thanks{tiago.araujol@ufrpe.br}
\affiliation{Departamento de F\'isica, Universidade Federal Rural de Pernambuco, Recife, PE 52171-900, Brazil}

\date{\today}

\begin{abstract}

The present work consists of a numerical study of the dynamics of irrational polygonal billiards. Our contribution reinforces the hypothesis that these systems could be Strongly Mixing, although never demonstrably chaotic, and discuss the role of rotational symmetries on the billiards boundaries. We introduce a biparametric polygonal billiards family with only $C_n$ rotational symmetries. Initially, we calculate for some integers values of $n$ the filling of the phase space through the Relative Measure $r(\ell, \theta; t)$ for a plane of parameters $\ell \times \theta$. From the resulting phase diagram, we could identify the completely ergodic systems. The numerical evidence that symmetrical polygonal billiards can be Strongly Mixing is obtained by calculating the Position Autocorrelation Function, $\Cor_x(t)$, these figures of merit result in power law-type decays $t^{- \sigma}$. The Strongly Mixing property is indicated by $\sigma = 1$. For odd small values of $n$, the exponent $\sigma \simeq 1$ is obtained while $\sigma < 1$, weakly mixing cases, for small even values. Intermediate $n$ values present $\sigma \simeq 1$ independent of parity. For high values of symmetry parameter $n$, the biprametric family tends to be a circular billiard (integrable case). This range shows even less ergodic behavior when $n$ increases and $\sigma$ decreases.


\end{abstract}

\maketitle


\section{Introduction}
\label{intro}

Ergodic theory is a branch of mathematics that classifies dynamical systems according to their degree of randomness. In ascending order: Ergodic ($E$), Mixing ($M$), Kolmogorov ($K$) and Bernoulli ($B$), being only systems $K$ and $B$ chaotic by having positive Kolmogorov-Sinai entropy $h_{\text{KS}}$ (or Lyapunov Exponent), so that $ E\supset M \supset K \supset B$ \cite{ozo:1988,ott:2002}. Inner the Mixing systems still exist a subclassification, the Weakly Mixing (WM) and Strongly Mixing (S) systems. Billiards are prototype models in the ergodic theory of Hamiltonian systems. Two-dimensional billiards correspond to a particle moving in a region with reflecting walls. Their dynamics can range from regular to completely chaotic, depending on the shape of the enclosure. Many results over the last decades provided a consistent path for a large amount of analytical and numerical work that boosted the field of non-linear dynamics \cite{str:2018}.

Nevertheless, some systems still need to be understood, particularly the polygonal billiards, which have $h_\text{KS}=0$, not chaotic. With a few exceptions, polygonal billiards are not integrable and exhibit random behavior, thus known as pseudo-integrable \cite{ric:1981}. These systems can be separated into two classes, the rational polygons, in which at least one internal angle is rational with $\pi$, and the irrational ones, all internal angles are irrational with $\pi$. On the dynamics of rational polygons, several mathematical results have been published over the last four decades, such as the work by Katok \cite{kat:1980}, which demonstrates that Strong Mixing behavior never occurs for this billiards class. Years later, \cite{ker:1986} dealt with ergodicity, already Gutkin and Katok \cite{gut:1988} proved the Weak Mixing related to polygons with vertical or horizontal sides, and recently this theme was revisited in \cite{sab:2017}. One of the last advances of such systems is the work of \'Avila and Delecroix \cite{avi:2016}, which deals with the connection between Weak Mixing and regular polygonal billiards. We emphasize that for the mathematical community, polygonal billiards are never Strongly Mixing \cite{gut:1996,che:2006}, but no proof of this fact has been given until now. At this point, to contribute to the understanding of irrational polygons, where about 20 years ago, some works \cite{art:1997}, particularly the one of Casati and Prosen (CP) \cite{cas:1999}, shed light on this question when they provided robust numerical evidence that irrational triangular billiards are Mixing. Years later, other work \cite{ara:2013} reinforced the CP hypothesis. Therefore, there is no robust study on the Mixing property in irrational polygonal billiards different from triangles. Only recently, \cite{ara:2021} has provided numerical evidence that irrational hexagons can be Mixing, corroborating the numerical evidence from CP. More specifically, a biparametric family of irrational polygonal billiards was introduced with the property of discrete rotation symmetry.

In Classical Mechanics, continuous symmetries lead to conserved quantities of the system. This result is a theorem due to the mathematician Emmy Noether \cite{noe:1918,jos:1998,lem:2007,lem:1993}. Discrete symmetries were studied only years later. M. Lutsky in \cite{lut:1979,lut:1981} introduced a method for deriving conserved quantities from discrete symmetries. M. Aguirre and J. Krause realized profound studies and explicitly generalizing and obtaining the point symmetry group, including in the covariant form \cite{agu:1991,agu:1991b,kra:1992}. G. Cicogna and G. Gaeta studied the presence of Lie point symmetries in dynamical systems either in Newton-Lagrange or Hamilton form \cite{cic:1992}. In Hamiltonian systems, properties of symmetry of the phase space come from an interplay between the symmetries of an integrable Hamiltonian and perturbations \cite{zas:2007}. In Quantum Mechanics, symmetries cause states of degenerate energy \cite{ley:1996,die:2005,men:2007,li:2020,die:2021,ara:2021,ara:2023,ara:2023b,zha:2023}.

Our results show that $C_n$-Symmetrical Polygonal Billiards (repeats itself under a rotation of $2\pi/n$) present the Strong Mixing property depending on the parity of the symmetry parameter $n$ for small $n$. The dependence with the parity is missed for intermediate values $n$. The impact of symmetry on ergodic properties has been discussed recently for the case of triangular billiards, with symmetry under reflection by a median \cite{zah:2022,zah:2023}. Thus, the present work also motivates the development of mathematical results of the ergodic theory of dynamical systems. This paper is organized as follows: in sec. \ref{pspace}, the constraints and parameters are introduced to obtain the geometric shape of the family of $C_n$-Symmetrical Polygonal Billiards. The general behavior of phase space is also presented. Sec. \ref{RM} presents an extensive numerical calculation of the relative measure for different symmetry parities. From some selected billiards, sec. \ref{cor} presents different correlation decays, indicating a weakly or strongly mixing behavior. Sec. \ref{high} shows how the results presented here are robust for high values of symmetry parameter $n$. The billiards family tends to a circular border for this range, wisely regular. Concluding remarks and perspectives are presented in the last section, besides a comment on the quantization of the proposed billiards.

\section{The $C_n$-Symmetrical Polygonal Billiards and Reduced Phase Space}
\label{pspace}

\begin{figure}[!htpb]
 \centering
 \includegraphics[width=0.9\columnwidth]{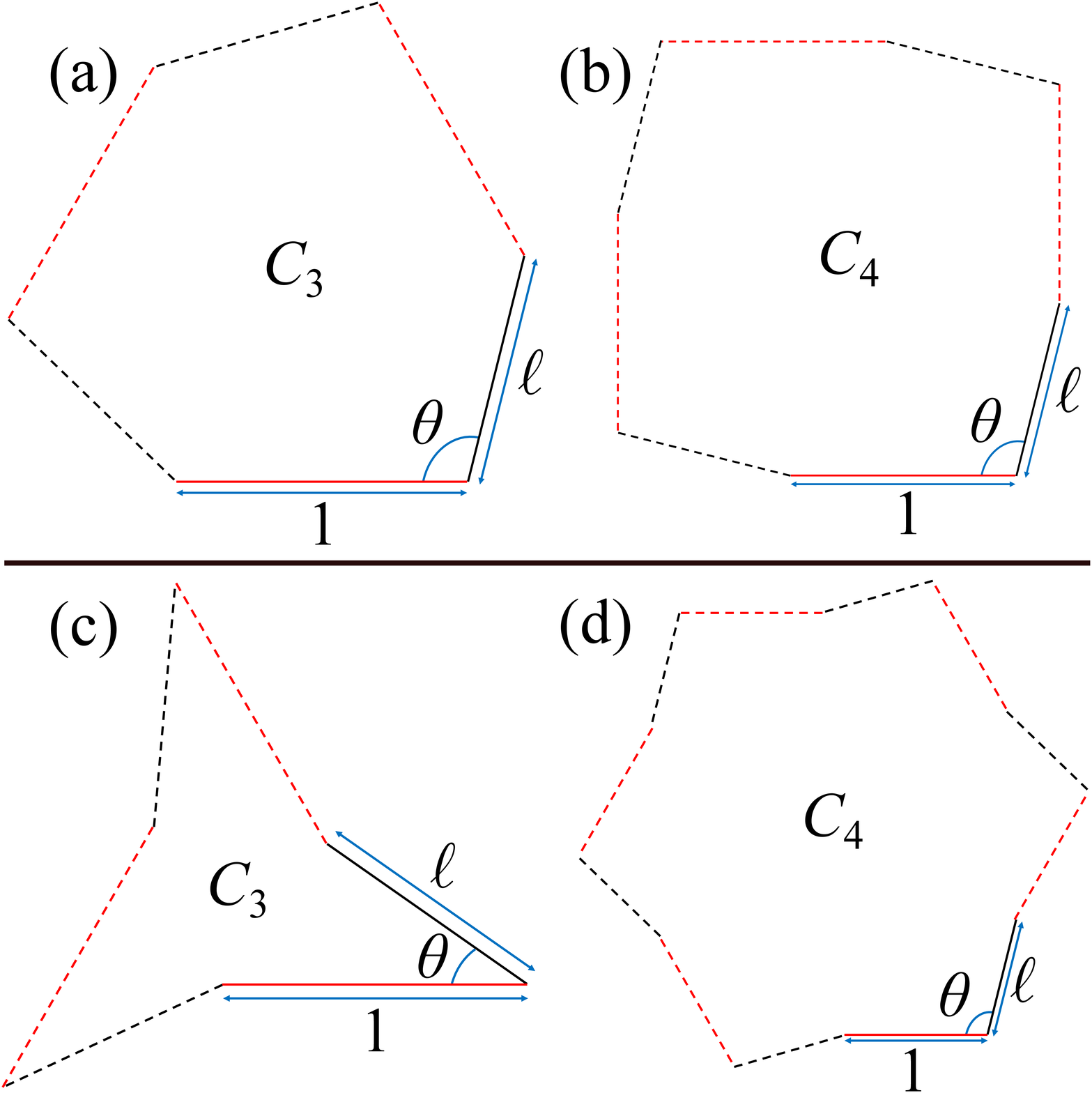}
 \caption{Examples of Symmetrical Polygonal Billiards with geometric parameters $(\ell, \theta)$. The boundaries are formed by alternating adjacent sides with unitary and $\ell$ lengths (red and black, respectively). These two sides form the angle $\theta$. (a) SPB with symmetry $C_3$. The boundary repeats itself after consecutive rotations of $120 \degree$ (dashed lines). (b) Another example of SPB, this $C_4$-symmetric case repeats the border after consecutive rotations of $90 \degree$ (dashed lines). (c)(d) Non-convex cases of the previous two are avoided here.}
 \label{fig:billiards}
\end{figure}

The billiards family introduced in this work are convex polygons of two alternating adjacent sides. One is with unit length, and the other is the segment $\ell \in (0,1)$. These two sides comprised an angle $\theta$. This pattern repeats $n$ times depending on the chosen symmetry, and the final shapes are symmetrical by rotation around the geometric center. The billiards are $C_n$-symmetric, for $n \geqslant 2$. Thus they have the same shape when rotated by an angle of $2\pi/n$. The total number of segments will be $2n$, and the resulting alternating angles are $\theta$ and $2\pi(1-1/n)-\theta$. The upper panels in Fig. \ref{fig:billiards} show two final generic shapes $C_3$-symmetric (on the left) and $C_4$-symmetric (on the right). The geometric construction so that the billiards do not become non-convex, as shown in the lower panels of Fig. \ref{fig:billiards}. This constraint requires that a chosen angle $\theta$ be greater than the minimum angle established by a given symmetry, namely, $\theta^{(\text{min})}_n=\pi(1-2/n)$, and its upper limit being $\theta^{(\text{max})}=\pi$. All billiard angles studied here are irrational with $\pi$, which the genus calculated are around $10^{17}$. From now on, we will use the abbreviation $C_n$-SPB to refer to \emph{$C_n$-Symmetrical Polygonal Billiards}.

\begin{figure}[!htpb]
 \centering
 \includegraphics[width=0.9\columnwidth]{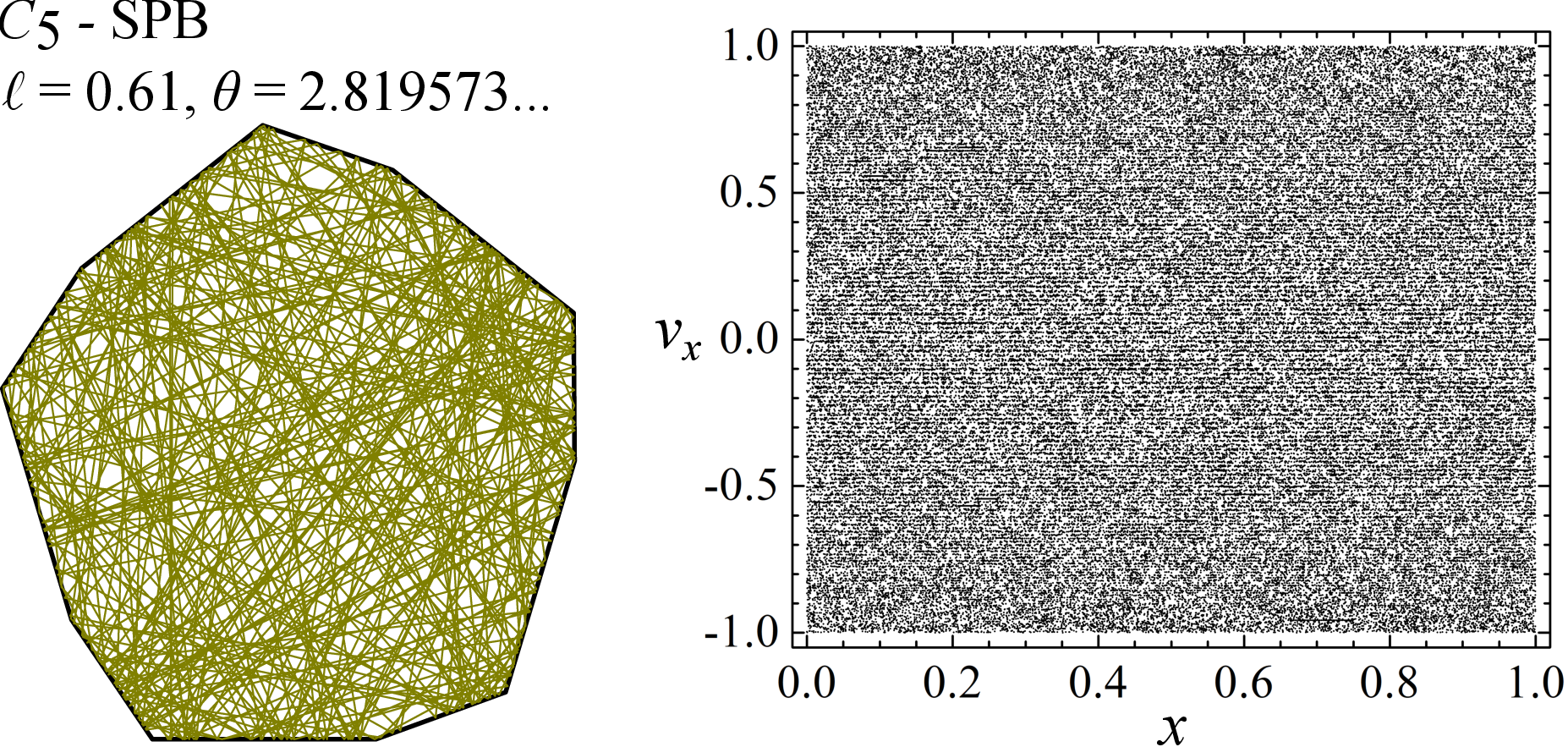}
 \caption{Left Panel: typical trajectory in the real space in the $C_5$-SPB after 200 impacts on the boundary. Right Panel: reduced phase space $(x,v_x)$ for a typical trajectory in the same billiard after $10^5$ collisions on the horizontal boundary.}
 \label{fig:pspace}
\end{figure}

As an example of dynamics in a real space in a $C_n$-SPB, the Left Panel of Fig. \ref{fig:pspace} shows a representation of 200 collisions for an arbitrary Initial Condition (IC) of a billiard $C_5$-SPB with $(\ell,\theta)=(0.61,2.819573...)$. The characterization of the dynamics of a given billiard will be carried out from collisions in a Poincar\'e section, i.e., although the entire boundary is part of the dynamics, we will only compute the interaction of the particle with a single segment, being the horizontal and of unitary length the chosen one \cite{cas:1999,ara:2013,ara:2015,ara:2021,ara:2023,ara:2023b}. In a discrete-time $t$, the particle departs from the section at position $x$ and a component velocity tangent to the border, defined by $v_x$. A reduced phase space is then defined by the intervals $0<x<1$ and $-1<v_x<1$. On the Right Panel of Fig. \ref{fig:pspace} is shown the reduced phase space with $t=10^5$ collisions for $C_5$-SPB with $(\ell,\theta)=(0.61,2.819573...)$, which it exhibits full ergodicity. The following section will investigate how the phase space filling evolves with time for different symmetries. Billiards with a fast tendency towards ergodicity will be candidates to be Strongly Mixing. 

\section{Relative Measure}
\label{RM}

In this section, we analyze how fast the filling of the phase space of a given billiard is dictated by its rotational symmetry. More specifically, the dependence with the parity of symmetry parameter $n$. For this, the reduced phase space $(x,v_x)$ is partitioned into a large number $N_\text{C}$ of cells. In our numerical calculations, we use $N_\text{C}=10^6$. Let $n(t)$ be the number of cells visited up to collision $t$ for a given trajectory, and $\av{n(t)}$ will be the average for different orbits with random ICs. Thus, the Relative Measure, i.e., the average fraction of visited cells, is $r(t) = \av{n(t)} /N_\text{C}$. As predicted for the Random Model (RM) \cite{rob:1997}, if all cells have the same probability of being visited, then $r(t) = r_{\text{RM}}(t)$, where
\begin{equation}
r_{\text{RM}}(t) = 1 - \textrm{exp} (-t/N_\text{C}).
\label{eq:RM}
\end{equation}

\begin{figure}[!htpb]
 \centering
 \includegraphics[width=0.9\columnwidth]{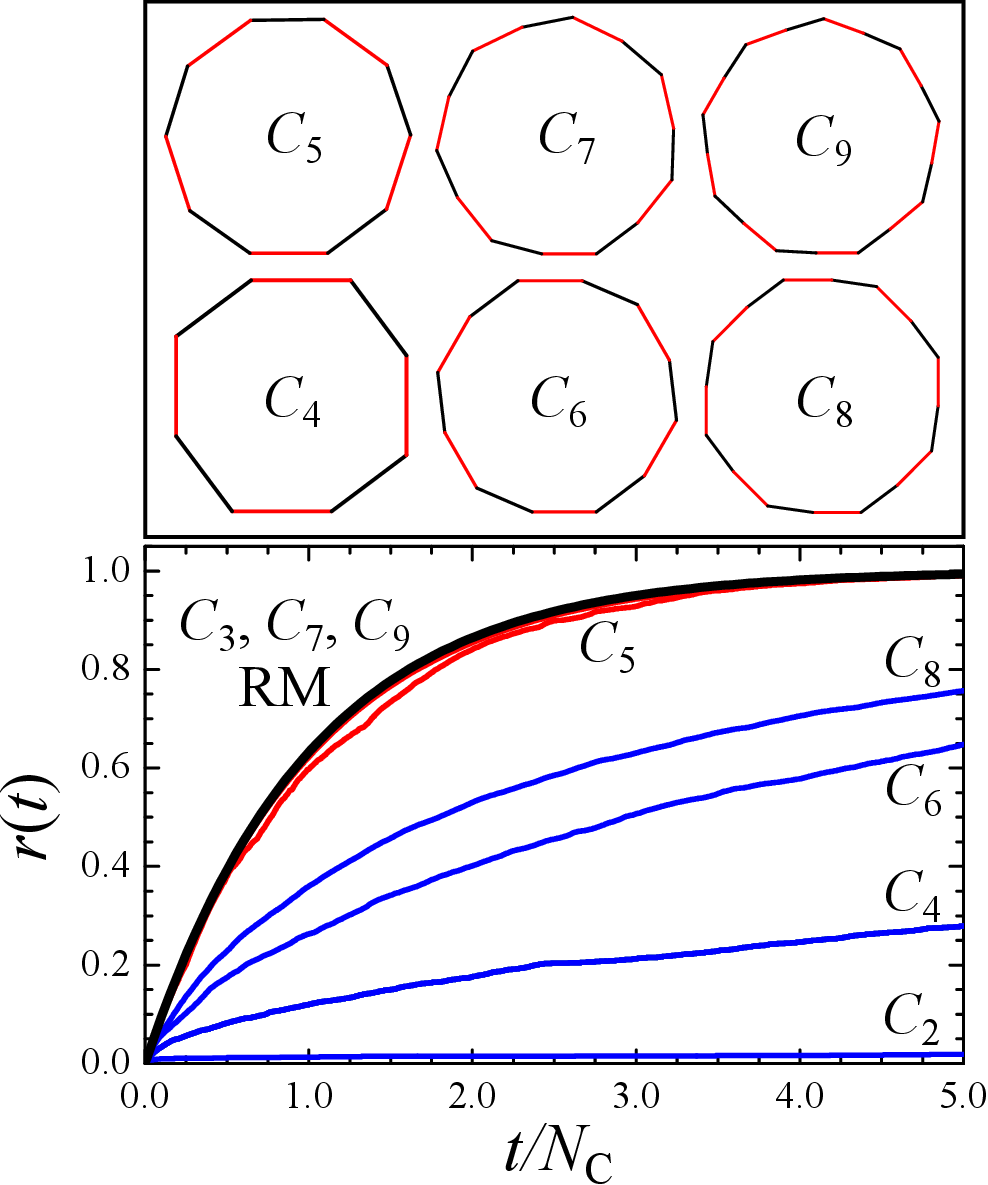}
 \caption{Upper Panels: $C_n$-SPB boundaries for some values of symmetry parameter $n$. Lower Panel: calculated Relative Measure $r(t) $for the same values of $n$. The geometric parameter values are $(\ell, \theta) = (0.94, 2.499721...)$ in all cases. The curves of RM, eq. (\ref{eq:RM}), $n=7$ and $n=9$ are almost indistinguishable. The $C_2$-SPB is a simple quadrilateral with low complex dynamics that almost does not scatter the phase space.}
 \label{fig:RM}
\end{figure}

In our first analysis, we will maintain the pair $(\ell, \theta) = (0.94, 2.499721...)$ while varying the value of symmetry parameter $n$. The Upper Panel in Fig. \ref{fig:RM} shows the borders of the resulting billiards ordered by the parity of $n$. The Relative Measure $r(t)$ for some symmetries is shown in the Lower Panel. The curves for even cases do not follow the result for eq. (\ref{eq:RM}) represented by the solid black line. The first integrant of the symmetrical family, the $C_2$-SPB, is a simple parallelogram of sides of length unitary and $\ell$ and angles $\theta$ and $2\pi-\theta$. This low complexity of the border leads to a phase space filled very slowly. On the other hand, $r(t)$ for odd symmetries follows the RM. So for this pair of parameters, we observe a fast filling of the phase space for the odd symmetries. Next, we will do a broader analysis over the entire range of the parameters $(\ell, \theta)$ as a function of the parity of the symmetries. 

\begin{figure}[!htpb]
 \centering
 \includegraphics[width=0.975\columnwidth]{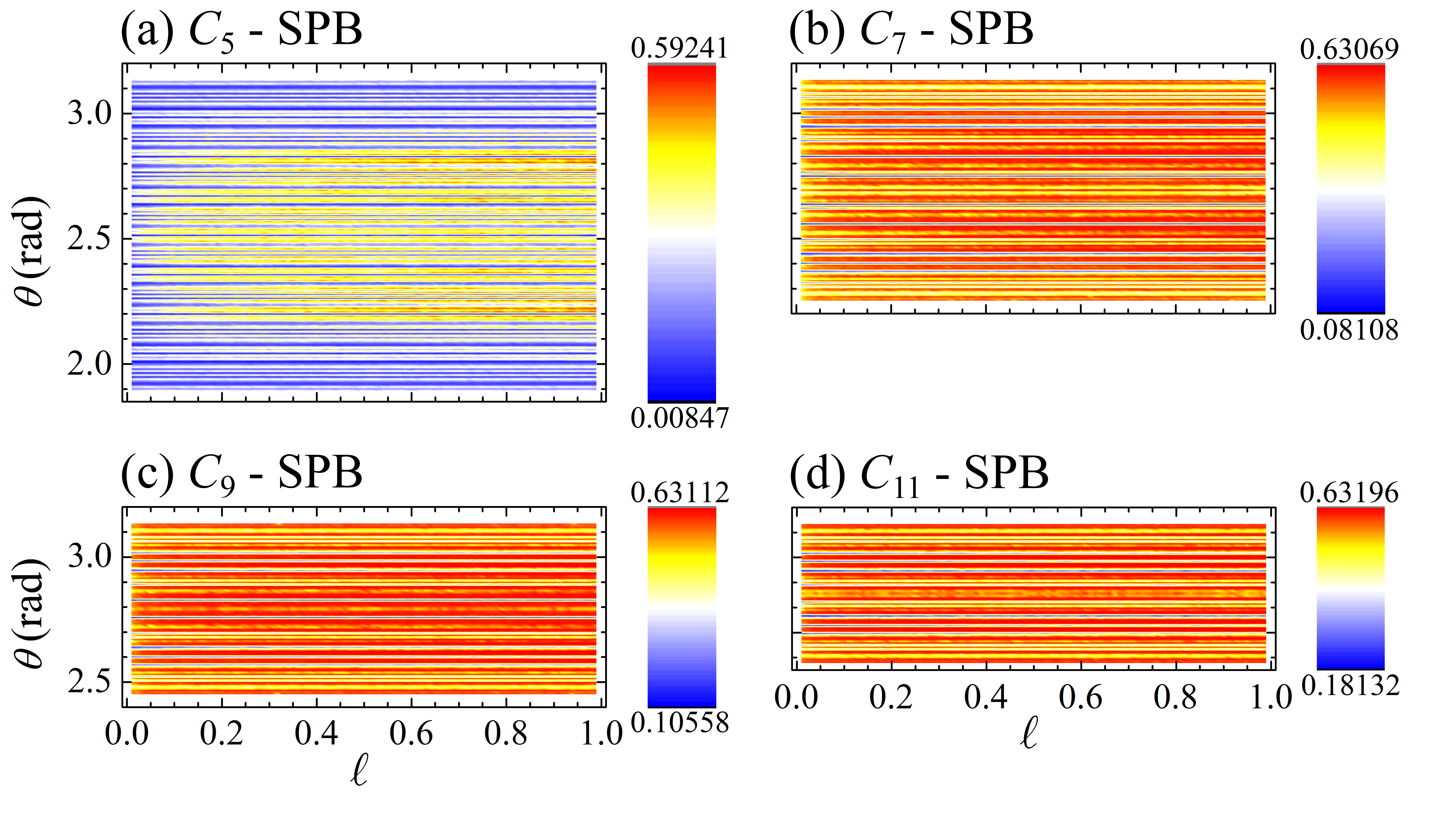}
 \caption{Phase diagram of $r(t=N_\text{C})$ on the parameter space $(\ell, \theta)$ for odd symmetries of $C_n$-SPB. For (RM), eq. (\ref{eq:RM}), $r_\text{RM}(t=N_\text{C}) \simeq 0.63$. (a) For $C_5$-SPB and $(\ell, \theta) = (0.65, 2.22267...)$, the dynamics presents a phase space filled almost like RM, resulting in $r(t=N_\text{C}) \simeq 0.59$. (b)(c)(d) $C_7$-SPB, $C_9$-SPB and $C_{11}$-SPB present a phase space filled like RM, resulting on $r(t=N_\text{C}) \simeq 0.63$ for $(\ell, \theta) = (0.64, 2.471759...)$, $(\ell, \theta) = (0.93, 2.64625...)$ and $(\ell, \theta) = (0.97, 2.71176...)$ respectively.}
 \label{fig:RM_odd}
\end{figure}

\begin{figure}[!htpb]
 \centering
 \includegraphics[width=0.975\columnwidth]{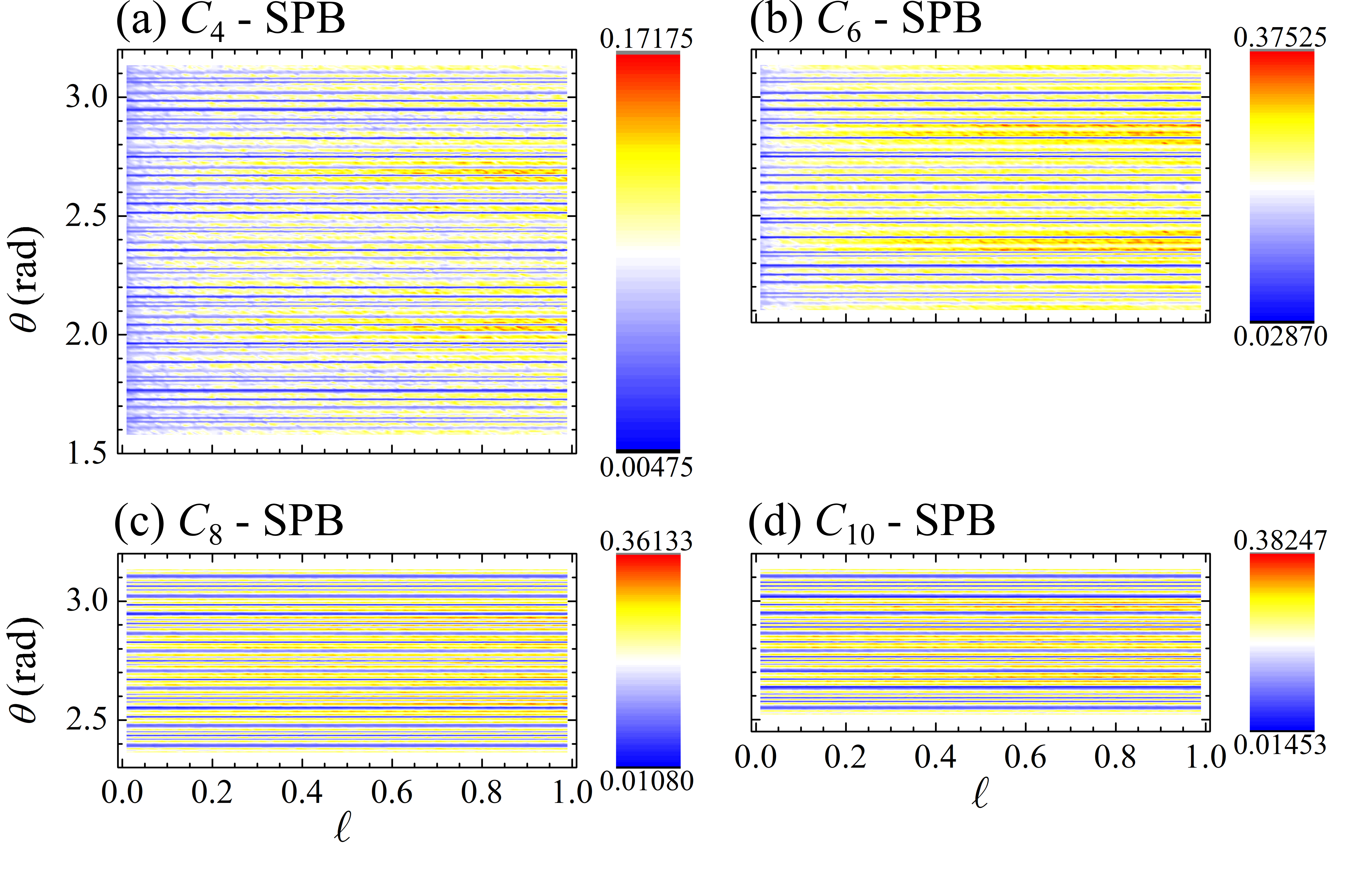}
 \caption{Phase diagram of $r(t=N_\text{C})$ on the parameter space $(\ell, \theta)$ for even symmetries of $C_n$-SPB. For (RM), eq. (\ref{eq:RM}), $r_\text{RM}(t=N_\text{C}) \simeq 0.63$. (a) $C_4$-SPB presents a phase space very few filled, resulting on $r(t=N_\text{C}) \simeq 0.17$ for $(\ell, \theta) = (0.88, 2.03418...)$. (b)(c)(d) $C_6$-SPB, $C_8$-SPB and $C_{10}$-SPB also present a phase space few filled, resulting in $r(t=N_\text{C}) \simeq 0.37$ for $(\ell, \theta) = (0.97, 2.35358...)$, $(\ell, \theta) = (0.99, 2.56825...)$ and $(\ell, \theta) = (0.8, 2.67821...)$ respectively. For small even values of $n$, this behavior remains due to many aligned parallel sides whose scatter trajectories are slowly in phase space.}
 \label{fig:RM_even}
\end{figure}

In order to look for billiards that may be strongly or weakly mixing, we must first look at the rate of ergodicity in such systems. The closer the behavior of $r(t)$ is to $r_{\text{RM}}$, the greater the chances of given billiards displaying the Strongly Mixing property. To map the ergodicity of the $C_n$-SPB family, we calculate $r(t=N_\text{C})$ for many billiards, up to 20,000, depending on the symmetry. For RM, $r_{\text{RM}}(t=N_\text{C}) = 0.632121...$ The results are shown in the phase diagrams in Figs. \ref{fig:RM_odd} and \ref{fig:RM_even}, separated into odd and even symmetries, respectively. Billiards that present $r(N_\text{C})$ close to $r_{\text{RM}}(N_\text{C})$ reached full ergodicity fastly. On the other hand, when comparing Figs. \ref{fig:RM_odd} and \ref{fig:RM_even}, we observe that even symmetries have a slower filling of phase space. For small even values of symmetry parameter $n$, this behavior remains due to many aligned parallel sides whose scatter trajectories are slowly in phase space. While for odd symmetries, the number of ergodic billiards is vast, highlighted in the red regions of Fig. \ref{fig:RM_odd}. Note that the range of the parameter $\theta$ decreases with $n$ increasing, and for each case of the phase diagrams, their values are symmetrical to the center of the range, $\theta^{(\text{mid})}_n=\pi(1-1/n)$.

\section{Decay of Correlations}
\label{cor}

\begin{figure}[!htpb]
 \centering
 \includegraphics[width=0.975\columnwidth]{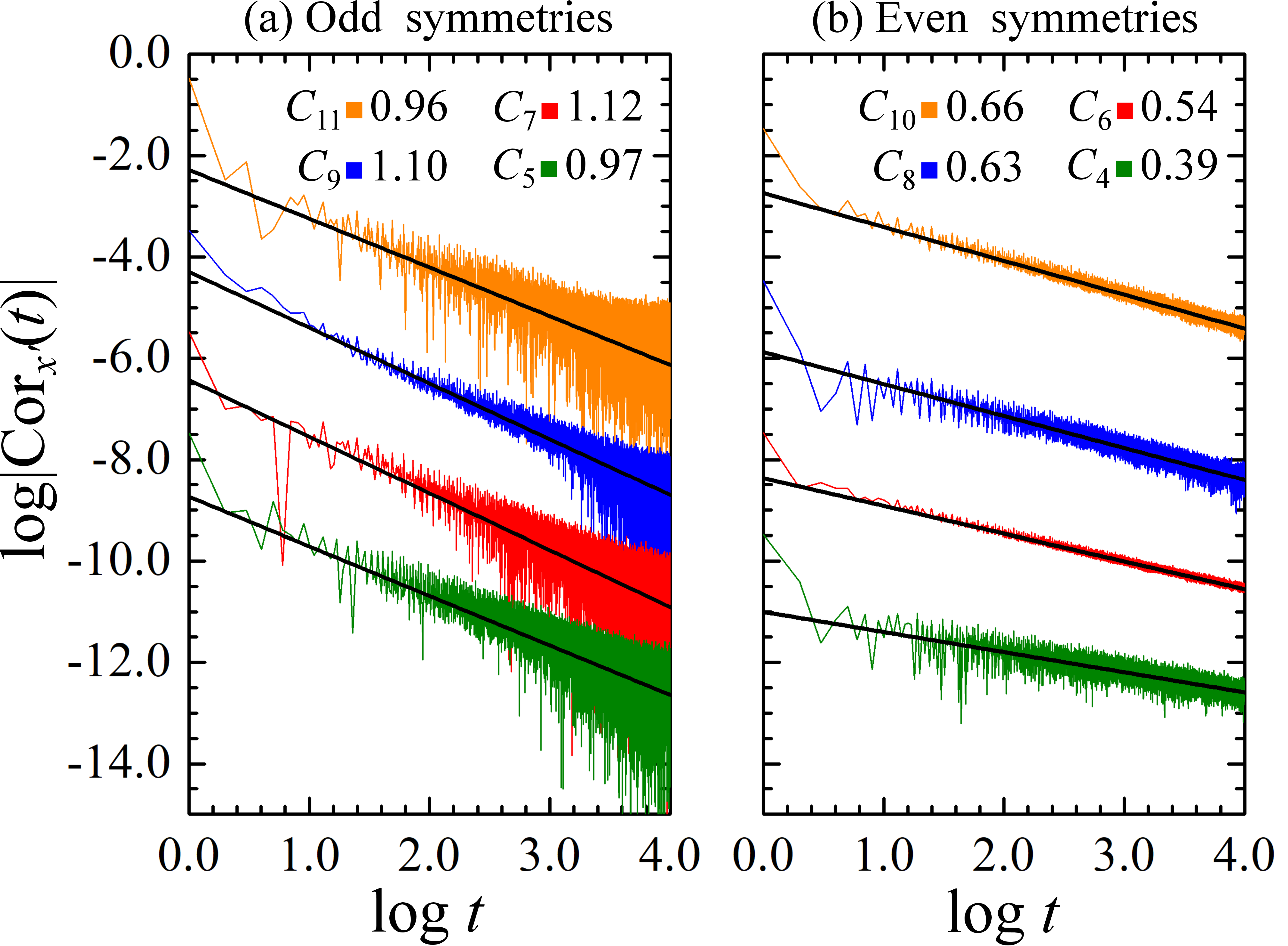}
 \caption{Calculated position autocorrelation function $\Cor_{x'}(t)$ in log-log scale for the $C_n$-SPB members with maximum $r(t=N_\text{C})$ from phase diagrams in Figs. \ref{fig:RM_odd} and \ref{fig:RM_even}. (a) Odd values of symmetry parameter $n$, from down to above, $C_5$, $C_7$, $C_9$ and $C_{11}$. The decays obey power laws $|\Cor_{x'}(t)| \sim t^{-\sigma}$ (black lines) with exponents $\sigma \geqslant 0.9$ (legend values), indicating Strongly Mixing dynamics. (b) Even values of $n$, from down to above, $C_4$, $C_6$, $C_8$ and $C_{10}$. The decays obey power laws (black lines) with exponents $\sigma < 1$ (legend values), indicating the Weakly Mixing property. All exponents were obtained from linear fits with errors of the order of 0.001.}
 \label{fig:cor_small}
\end{figure}

The characterization of the mixing dynamics of billiards is performed from the time-averaged position autocorrelation function,
\begin{equation}
\Cor_x(t) = \lim_{T \rightarrow \infty} \dfrac{1}{T} \sum_{\tau=0}^{T-1}  x(\tau) x(\tau + t) - \av{x}^2.
\label{eq:cor}
\end{equation}
CP showed numerical evidence, through these functions, that the irrational triangular billiards are mixing \cite{cas:1999}. This hypothesis was strengthened later with an extensive investigation into a wider variety of triangles and hexagons \cite{ara:2013,ara:2021}. The classification of the Mixing behavior is dictated by the power law decay of the autocorrelation function, $|\Cor_x(t)| \sim t^{-\sigma}$. When $\sigma \simeq 1$, there is numerical evidence that the system is Strongly Mixing. Next, we will analyze the dynamic behavior of billiards that presented the highest values $r(t=N_\text{C})$ shown in the phase diagrams of Figures. \ref{fig:RM_odd} and \ref{fig:RM_even}. Thus, some of these can present a fast filling of the phase space and consequently become candidates to be Strongly Mixing. We perform the calculations in a rescaled position $x'=2x-1$ so that the term $\av{x'}$ can be neglected in eq. (\ref{eq:cor}), since $\av{x} \simeq 0.5$. All tested cases present $\av{x'} \simeq 10^{-8}$. For odd symmetries ($C_5$ to $C_{11}$), the Left Panel of Figure \ref{fig:cor_small} shows their autocorrelation function in a log-log scale. A tendency of fast decay of the autocorrelations is observed, with $\sigma \geqslant 0.9$, emphasizing the billiards with symmetries $C_7$ and $C_9$ presenting $\sigma \simeq 1$, being numerical evidence of a strongly mixing dynamics for these systems. All exponents were obtained from fits with errors of the order of 0.001. The curves have been shifted downwards for better visualization. For even symmetries ($C_4$ to $C_{10}$), $\Cor_{x'}(t)$ calculations are shown in the right panel of Figure \ref{fig:cor_small}, with all exponents $\sigma < 1$ providing evidence of Weakly Mixing dynamics. As in the odd cases, all exponents were obtained from fits with errors around 0.001, and the curves shifted downwards. In the next section, we will analyze the dynamics of billiards in symmetries superior to those already discussed.

\section{Higher Symmetries}
\label{high}

\begin{figure}[!htpb]
 \centering
 \includegraphics[width=0.975\columnwidth]{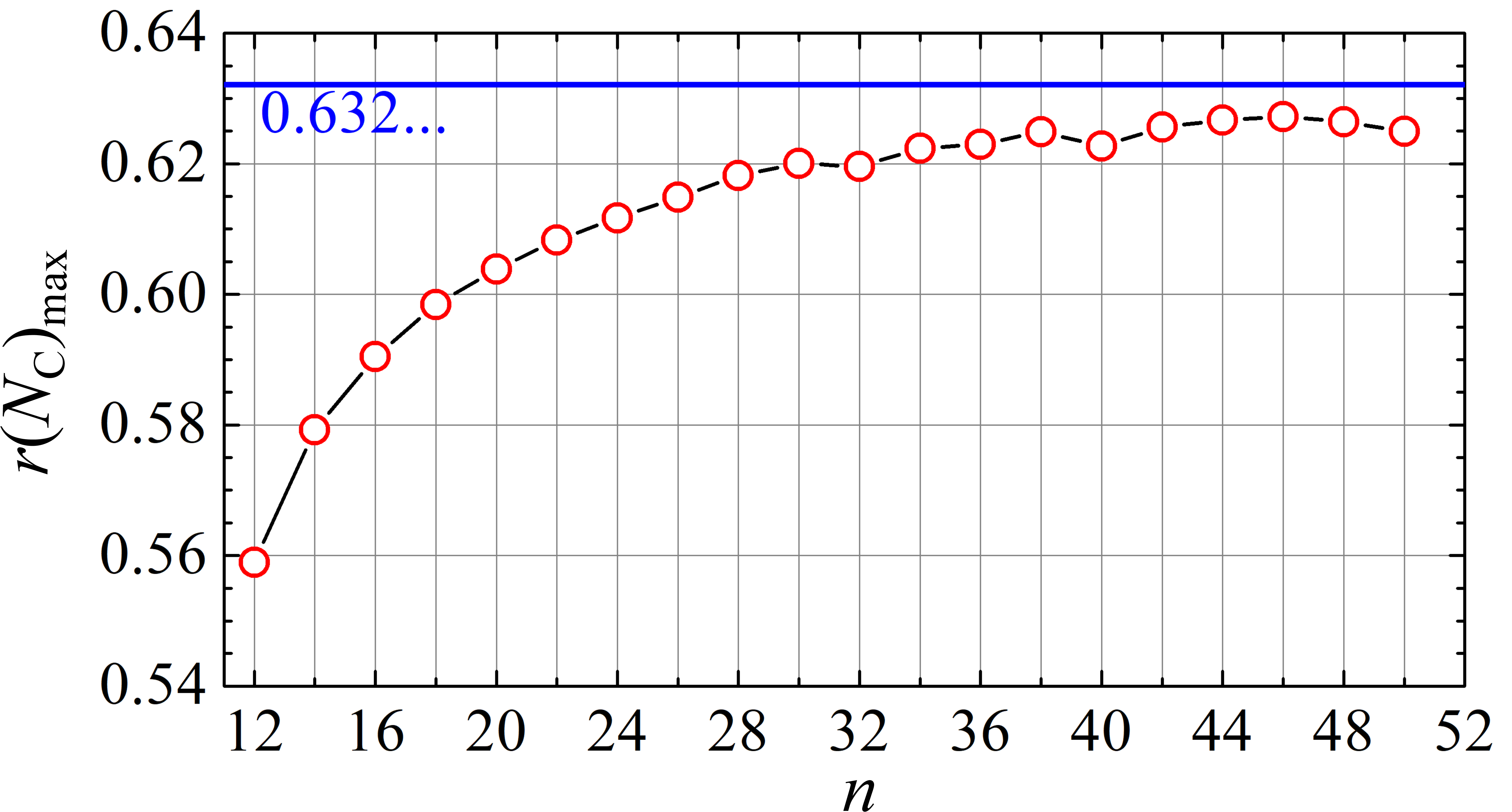}
 \caption{Maximum value of $r(N_\text{C})$ for $\ell=0.8$ for even values of the symmetry parameter $n$. For small symmetry values, $r(N_\text{C})_\text{max}$ present values far from 0.632..., the expected for the RM. While for major values of $n$, $r(N_\text{C})_\text{max}$ towards 0.632..., indicating the possibility of reaching strongly mixing dynamics for billiards with these parameters.}
 \label{fig:MR_odd_even}
\end{figure}

\begin{figure}[!htpb]
 \centering
 \includegraphics[width=0.975\columnwidth]{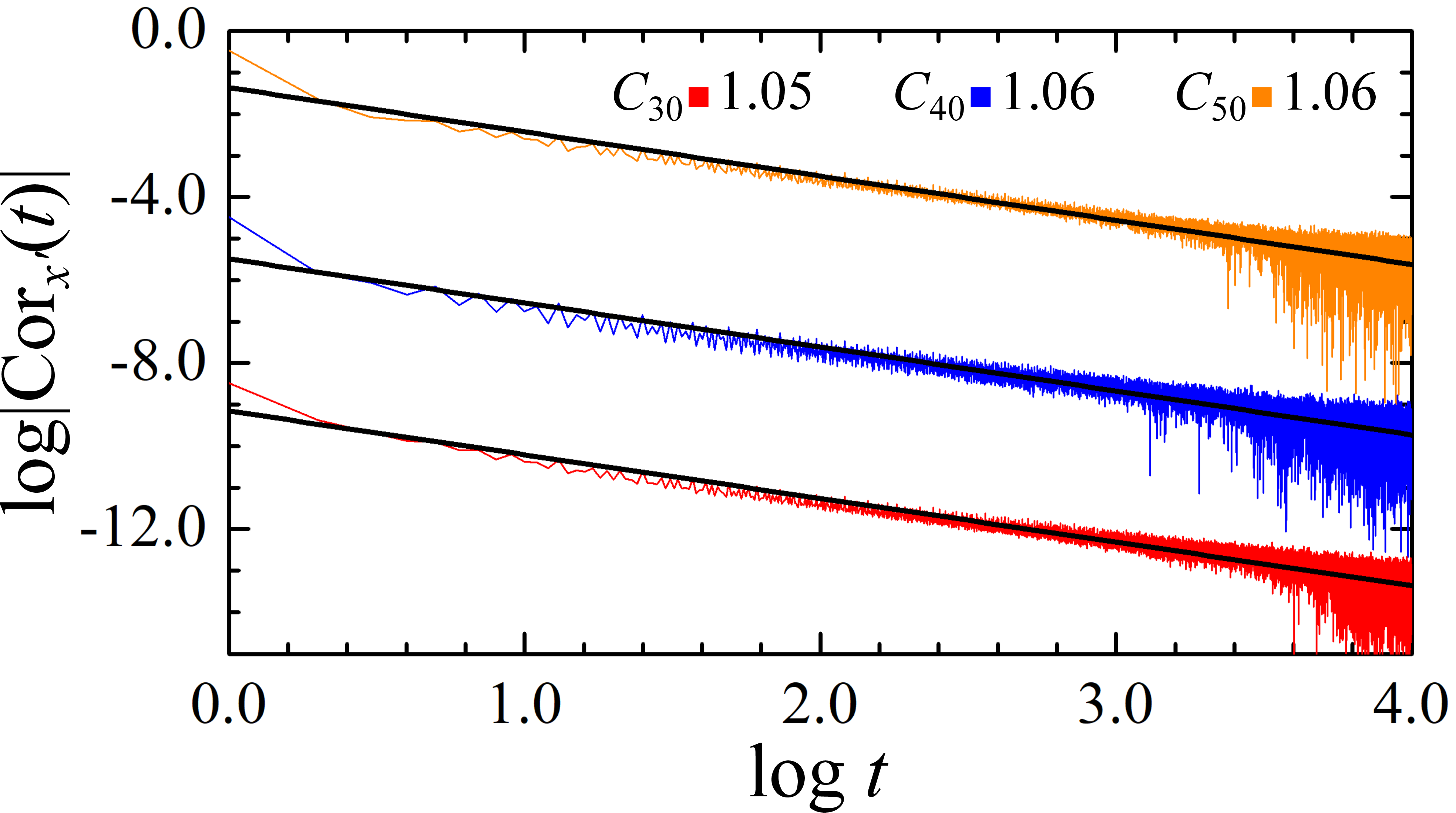}
 \caption{Calculated position autocorrelation function $\Cor_{x'}(t)$ in log-log scale for the $C_n$-SPB members with maximum $r(t=N_\text{C})$ from the results in Fig. \ref{fig:MR_odd_even}. Even values of $n$, from down to up, $C_{30}$ ($\theta = 2.98634...$), $C_{40}$ ($\theta = 3.0261...$), and $C_{50}$ ($\theta = 3.04655...$). $\ell=0.8$ in all cases. The decays obey power laws $|\Cor_{x'}(t)| \sim t^{-\sigma}$ (black lines) with exponents $\sigma \simeq 1$ (legend values), indicating Strongly Mixing dynamics. All exponents were obtained from linear fits with errors of the order of 0.001.}
 \label{fig:cor_high1}
\end{figure}

For small values of the symmetry parameter $n$, the parity plays a fundamental role in the dynamics of $C_n$-SPB. All investigations for odd small $n$ present the possibility of reaching Strongly Mixing dynamics, while all even small $n$ do not present this possibility, being observed just weakly mixing dynamics. However, this pattern does not occur for any value of $n$. When $n$ increases, the parity significance is lost. To investigate major even values of symmetry parameters, we build up a simplified version of plane phases in Figure \ref{fig:RM_even}. In their, $r(t=N_\text{C})$ is approximately constant along the parameter $\ell$ so that we will set $\ell=0.8$ and vary $\theta$ searching for the maximum value of $r(t=N_\text{C})$. These results are summarized in Figure \ref{fig:MR_odd_even}, which shows that when $n$ increases, $r(N_\text{C})_\text{max}$ also increases towards the value of the RM (solid blue line). This major filling indicates the possibility of strongly mixing dynamics for billiards with these parameters, Figure \ref{fig:cor_high1} confirms this chance. A tendency of fast decay of the autocorrelations is observed, with $\sigma \geqslant 0.9$, emphasizing the billiards with symmetries $C_{30}$, $C_{40}$ and $C_{50}$ (parameters in figure label) presenting $\sigma \simeq 1$. All exponents were obtained from fits with errors of the order of 0.001. The curves have been shifted downwards for better visualization. This phenomenon occurs due to the complexity of boundaries when $n$ increases. For small symmetry parameters in secs. \ref{RM} and \ref{cor} the billiards are polygons with relatively few sides. The number of sides dictates how the velocities are scattered through the particle dynamics.

\begin{figure}[!htpb]
 \centering
 \includegraphics[width=0.9\columnwidth]{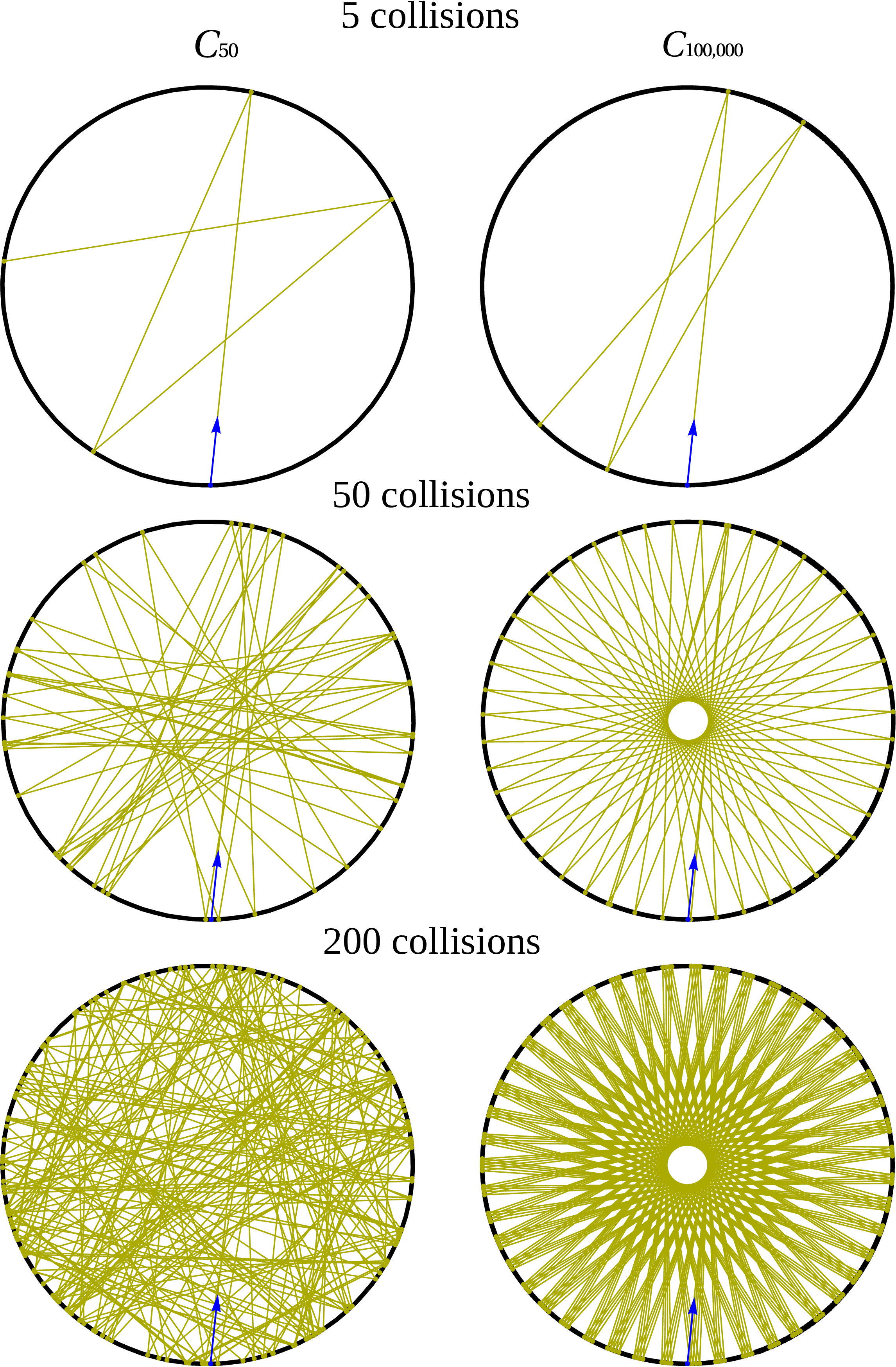}
 \caption{Real trajectories for $C_{50}$ ($\theta = 3.04655...$, $\ell=0.8$) (left) and $C_{100,000}$ ($\theta = 3.141561...$, $\ell=0.5$) (right) for some number of collisions from the same IC $(q/L,p)=(3\cdot10^{-6}, 0.1)$, the blue arrow indicates the IC. Upper Panel: real trajectory for 5 collisions, a few deviations between two billiards. Middle Panel: real trajectory for 50 collisions, the deviation between two billiards becomes relevant. Bottom Panel: real trajectory for 200 collisions, the deviation becomes evident. The real trajectory is scattered for $C_{50}$-SPB. While for $C_{100,000}$-SPB, the trajectory seems regular as in a circular billiard (see Fig. \ref{fig:C_high}). A dynamical visualization can be seen in a movie in supplemental material \cite{car:2023}.}
 \label{fig:traj}
\end{figure}

\begin{figure}[!htpb]
 \centering
 \includegraphics[width=0.975\columnwidth]{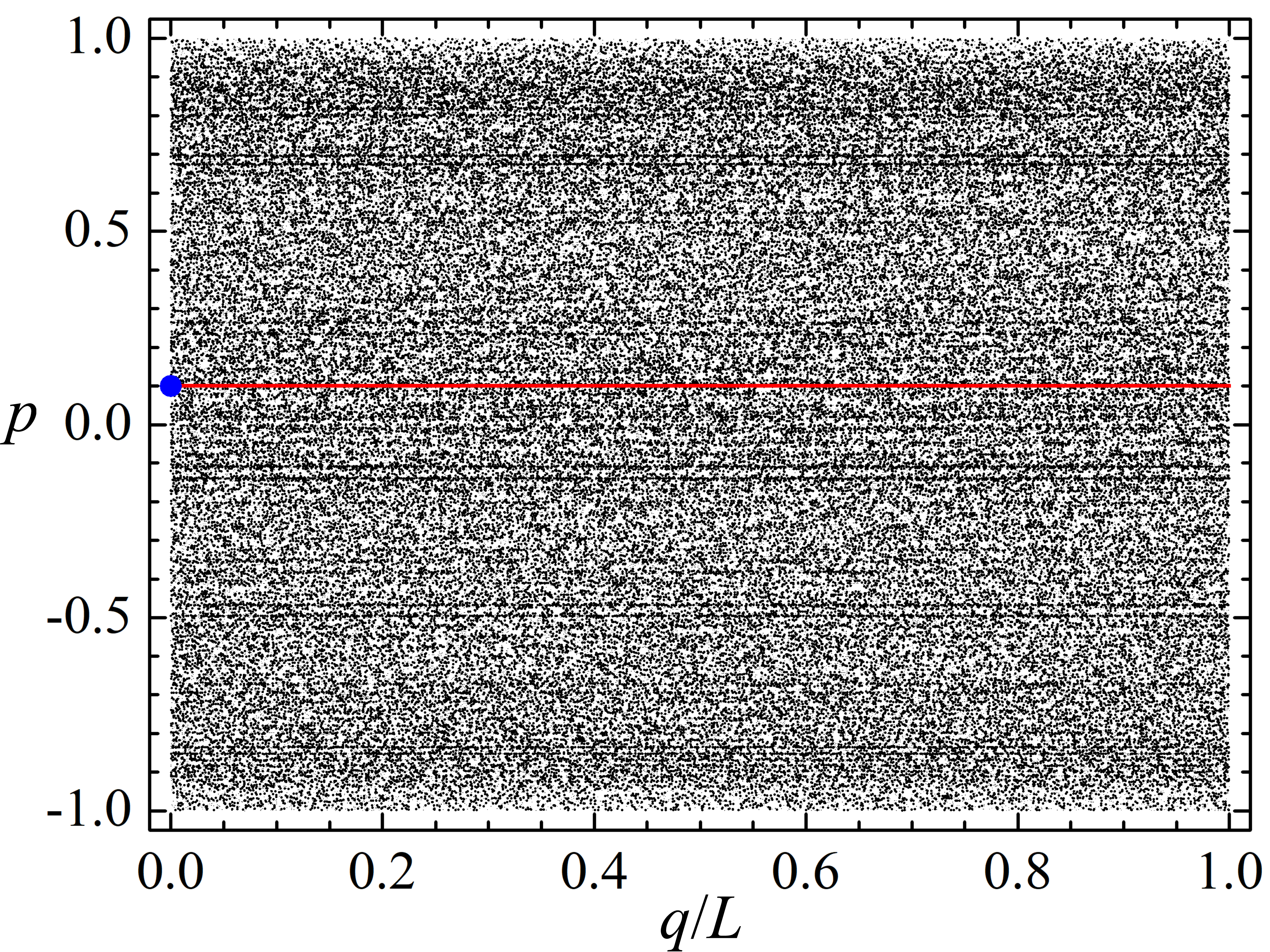}
 \caption{
 Phase spaces for $C_{50}$-SPB (black dots) and $C_{100,000}$-SPB (red dots) for $10^5$ colisions from the same IC $(q/L,p)=(3\cdot10^{-6}, 0.1)$ (blue point). As the real trajectory (see Fig. \ref{fig:traj}), the phase space for $C_{50}$-SPB is scattered. While for $C_{100,000}$-SPB, the phase space seems regular as in a circular billiard, with fixed $p$. The deviation in $p$ after $10^5$ collisions is of the order of $10^{-4}$.}
 \label{fig:delta_PS}
\end{figure}

\begin{figure}[!htpb]
 \centering
 \includegraphics[width=0.975\columnwidth]{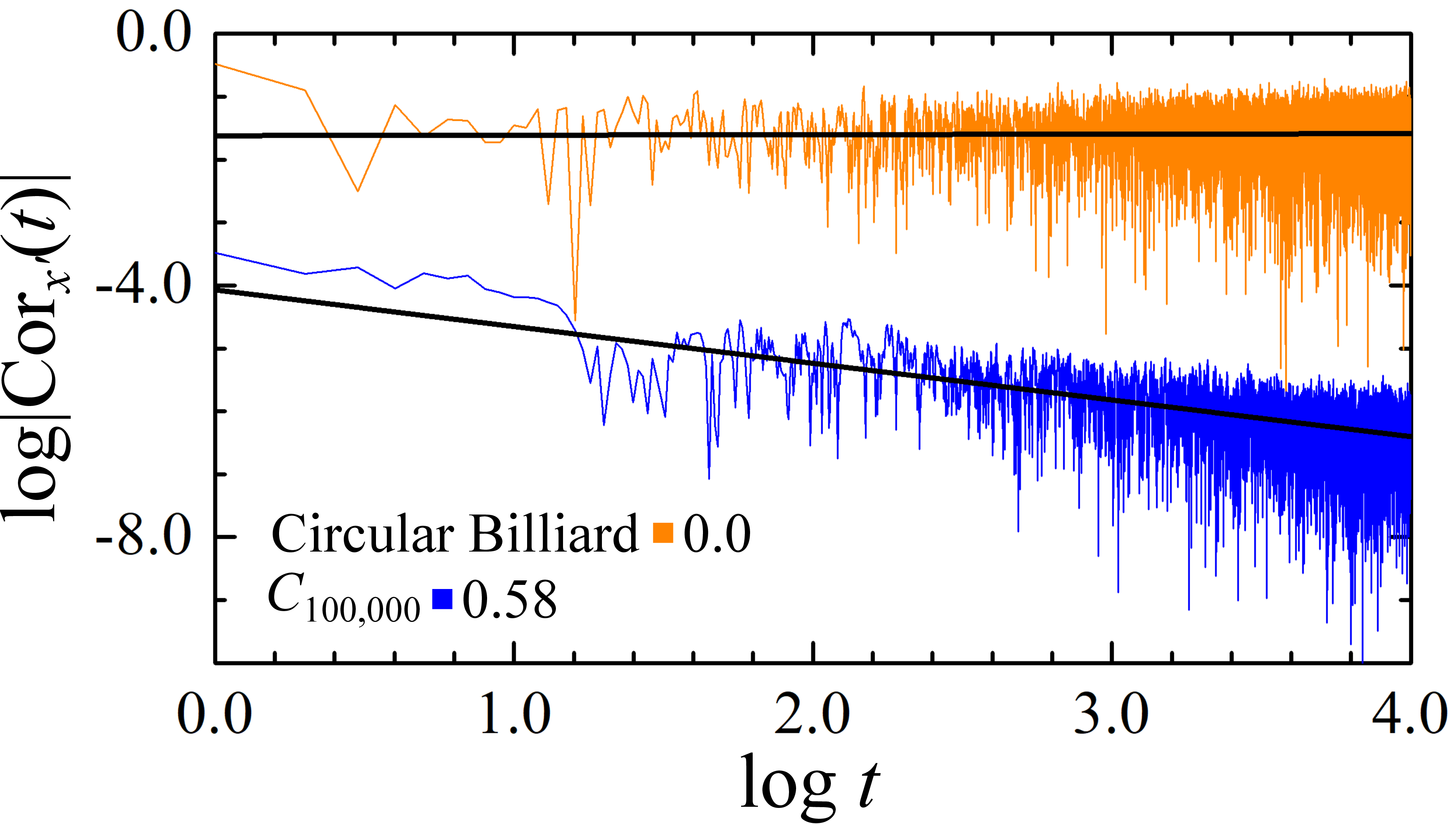}
 \caption{Calculated position autocorrelation function $\Cor_{q'}(t)$ in log-log scale for $C_{100,000}$-SPB and the Circular Billiard, the integrable case in eq. (\ref{eq:circ}). The decays obey power laws $|\Cor_{q'}(t)| \sim t^{-\sigma}$ (black lines) with exponent $\sigma \simeq 0.58$ for $n=$100,000. For the Circular Billiard, the correlation function oscillates with no decay behavior, $\sigma \simeq 0$. All exponents were obtained from linear fits with an error of around 0.01.}
 \label{fig:C_high}
\end{figure}

The $C_n$-SPB tends asymptotically to circular billiards (non-ergodic) when $n \gg 1$. The resulting shape is a polygon comprised of numerous edges forming almost shallow angles. This part of the text aims to analyze how robust is the numerical formation of these billiards and if the regular behavior of a circular billiard can be accessed for some finite value of $n$. To evolve these questions, we need to give up the reduced phase space because there is no flat side in a Circular Billiard, just a point where the tangent to the curve is horizontal. So we perform our analysis in the canonical Birchoff coordinates $(q,p)$. Where $q$ is the fraction of the perimeter, and $p$ is the tangent velocity of the border on a collision. So that $0 < q < L$, where $L$ is the billiard perimeter and $-1 < p < 1$. In this frame, the Circular Billiard has an analytical map of discrete time $t$ \cite{che:2006}, for unitary perimeter:
\begin{equation}
 \left\{ \begin{array}{l}
  q_t= q_0 + t \left ( \pi - 2\arcsin p_0 \right )/L \quad \mod 1\\
  p_t = p_0. \end{array}\right.
 \label{eq:circ}
\end{equation}
The most scattered trajectory possible for this map has caustics around the circle's center. Figure \ref{fig:traj} shows the trajectories in the real space in two $C_n$-SPB for $n=50$ ($\theta = 3.04655...$, $\ell=0.8$) (left panel) and $n=$100,000 ($\theta = 3.141561...$, $\ell=0.5$) (right panel) from the same IC $(q/L,p)=(3\cdot10^{-6}, 0.1)$. Throughout dynamics, the deviation of the trajectory from the expected for $n\rightarrow \infty$ is prominent for $n=50$ while for $n=$100,000 the fixed value of $p$ deviates merely $\sim 10^{-4}$ after $10^5$ collisions. A dynamical visualization of 200 collisions can be seen in the supplemental material \cite{car:2023}. The phase space in Birchoff coordinates after $10^5$ collisions, shown in the right Panel of Figure \ref{fig:delta_PS}, presents a variety of accessed values for $n=50$, $r(N_\text{C}) \simeq 0.625$. While for $n=$100,000, the unique value of $p$ acessed results in $r(N_\text{C}) = 0.001$. In Figure \ref{fig:C_high}, the autocorrelation functions for $n=$100,000, $\Cor_{q'}(t)$ dacays with exponent $\sigma \simeq 0.58$ indicating the lost of Strongly Mixing property, keeping the Weakly Mixing. For the regular dynamics in a Circular Billiards is expected a oscilation, the resulting $|\Cor_{q'}(t)|$ presents $\sigma \simeq 0$. The $C_n$-SPB family only assumes the regular behavior, as a Circular Billiard, for $n\rightarrow \infty$.
%
%
%
%

\section{Conclusions ans Perspectives}
\label{conc}

This paper presents numerical results on classical dynamics in a symmetric irrational polygonal billiards family. The $C_n$-SPB are convex polygons of two alternating sides with length unitary and $\ell \in (0,1)$. These sides form an angle $\theta$, and the resulting boundary is symmetric under rotations by $2\pi/n$. We investigate the possibility of the classical dynamics of these billiards being Strongly Mixing in the Ergodic Hierarchy of Hamiltonian Systems. We start exploring if, for some set of geometrical parameters $(\ell,\theta)$ and the symmetry parameter $n$, the respective dynamics tend to obey the Random Model (RM), eq. (\ref{eq:RM}) \cite{rob:1997}. We observe that for the initial symmetry parameters, $2 \leqslant n \lesssim 12$, the parity of $n$ plays a fundamental role in the behavior of the phase space. For odd $n$, the more scattered dynamics found are very near RM (Figure \ref{fig:RM_odd}). While for even $n$, the more scattered phase spaces are far from RM (Figure \ref{fig:RM_even}). The evidence of ergodicity is extended to the Mixing property by calculating the autocorrelation functions $\Cor_{x'}(t)$, eq. (\ref{eq:cor}). For selected parameters from previous analyses and odd symmetries, we observe a tendency of fast decay of $\Cor_{x'}(t)$, emphasizing that these billiards can present Strongly Mixing Dynamics (Figure \ref{fig:cor_small}). While for even values of the symmetry parameter $n$, just the Weakly Mixing dynamics can be reached (Figure \ref{fig:cor_small}).

The relevance of the parity of the symmetry parameter $n$ for the dynamics of $C_n$-SPB does not always occur. When $n$ increases the filling of the phase spaces towards RM even for even $n$ (Figure \ref{fig:MR_odd_even}). Thus, the Strongly Mixing dynamics also could be reached for even values of the symmetry parameter in this intermediary range of $n$. This effect occurs due to the complexity of boundaries when $n$ increases. For small symmetry parameters, the billiards are polygons with relatively few sides. The number of sides dictates how the velocities are scattered through the particle dynamics. The $C_n$-SPB tends asymptotically to circular billiards (non-ergodic) when $n \gg 1$. Figures \ref{fig:traj} and \ref{fig:delta_PS} show how the dynamics are sensitive to the polygonal boundaries. The Strongly Mixing property is lost for $n \gg 1$, the autocorrelation functions decay with exponent $\sigma < 1$ indicating a Weaky Mixing dynamics (Figure \ref{fig:C_high}). The $C_n$-SPB family only assumes the regular behavior, as a Circular Billiard ($\sigma = 0$), for $n\rightarrow \infty$.

\begin{figure}[!htpb]
 \centering
 \includegraphics[width=0.975\columnwidth]{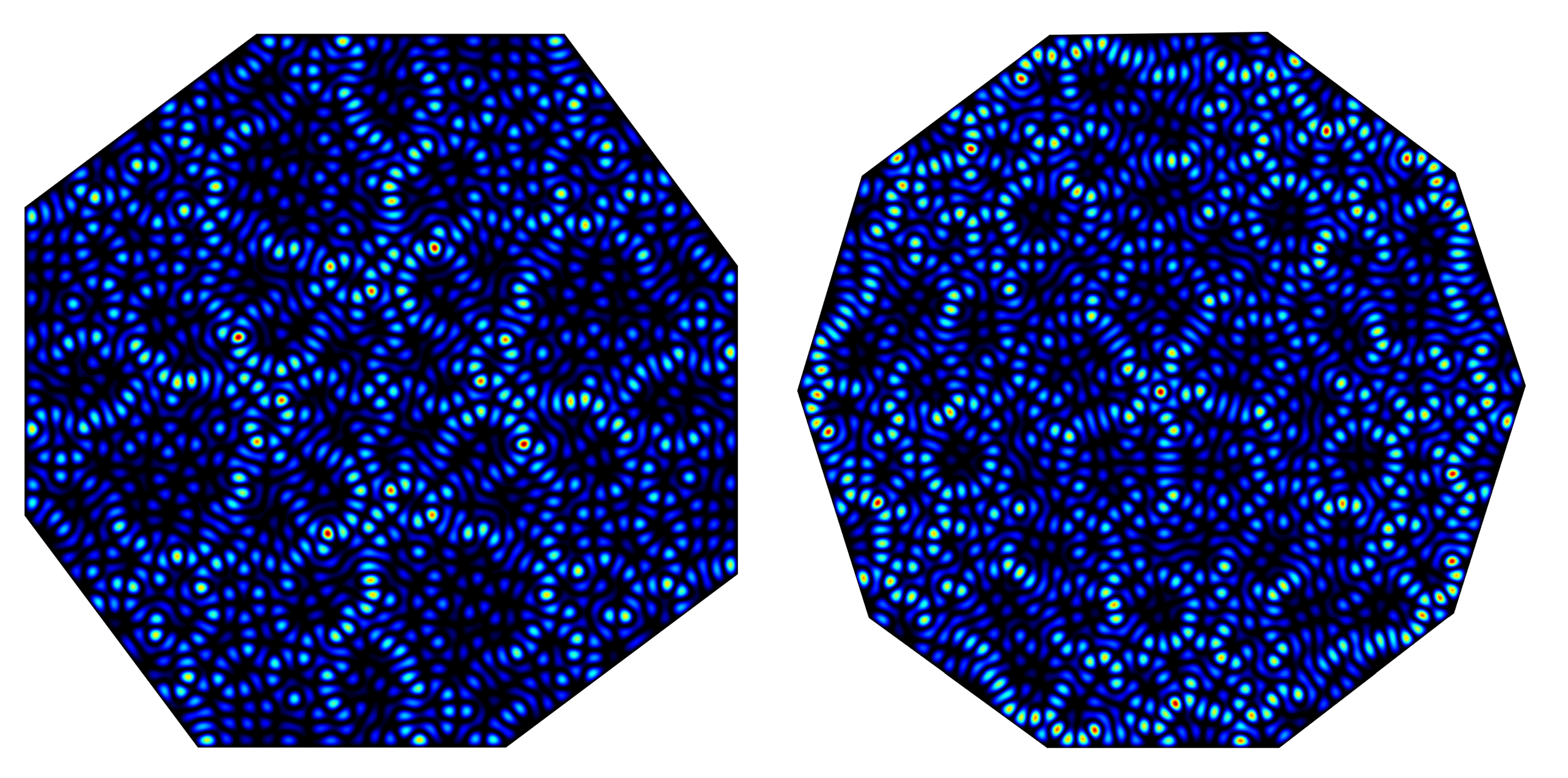}
 \caption{Probability density $|\psi(\vec{r})|^2$ obtained numerically from the Helmholtz equation using VS method \cite{ver:1995}. Small red tones indicate high probability values, while regions with dark tones indicate low values. Left Panel: around $2,000$th level for $C_4$-SPB and $(\ell,\theta) = (0.94,2.499721...)$. Right Panel: same for $C_5$-SPB.}
 \label{fig:quantum}
\end{figure}

As a perspective for future work, quantizing $C_n$-SPB is a meritorious investigation. The quantum properties of pseudo-integrable systems have been studied over the past decades \cite{ric:1981,lib:1994}. However, irrational billiards have yet to be investigated \cite{men:2007,ara:2013,ara:2021}. Considering that the discrete rotational symmetries in quantized billiards produce independent spectra of singlets and doublets (degenerate states) \cite{ley:1996,li:2020}, it should be explored how the spectral statistics are affected in the $C_n$-SPB, and how they are related to their classical counterparts (from weakly mixing to strongly mixing). The case of $n=3$ was explored in irrational hexagons \cite{ara:2021}, and their spectra were analyzed with intermediate formulas between Poisson and Random Matrices Theory statistics \cite{sto:2000,meh:2004}. Such formulas should be tested for $n>3$. Furthermore, the superposition of independent spectra could be studied as proposed by \cite{abu:2009}. The eigenfunctions associated with the singlet states are symmetrical with respect to the center of the billiards, as Figure \ref{fig:quantum} shows the probability density $|\psi(\vec{r})|^2$ of billiards with $C_4$ and $C_5$ symmetries. They were obtained by solving the Helmholtz equation, $\bigtriangledown^2 \psi_i(\vec{r}) = -k_i^2 \psi_i(\vec{r})$, with Dirichlet boundary conditions using a scaling method \cite{ver:1995}. Where $k_i^2$ is an energy eigenvalue. Aspects related to the intensity distribution of the eigenfunctions associated with singlets and doublets must be studied in depth \cite{mul:1997,mar:2012}.

\begin{acknowledgments}

Valuable discussions with F. M. de Aguiar are gratefully acknowledged. We are also grateful for the computational resources of LaSCoU and LCR from the Department of Physics at Universidade Federal Rural de Pernambuco. The Brazilian Agencies CNPq, CAPES, and FACEPE have supported this work.

\end{acknowledgments}


\bibliography{refs_2023_RBdoCarmoTAraujoL_SymmetricalBilliards.bib}

\end{document}